# On modifying properties of polymeric melts by nanoscopic particles

**Canan Atilgan,\* Ibrahim Inanc, Ali Rana Atilgan**

Sabanci University, Faculty of Engineering and Natural Sciences, Tuzla Istanbul 34956 Turkey

\*Corresponding author

*e-mail*: canan@sabanciuniv.edu

*telephone*: +90 (216) 4839523

*telefax*: +90 (216) 4839550

**ABSTRACT**

We study geometric and energetic factors that partake in modifying properties of polymeric melts via inserting well-dispersed nanoscopic particles (NP). Model systems are polybutadiene melts including 10–150 atom atomic clusters (0.1–1.5% v/v). We tune interactions between chains and particle by van der Waals terms. Using molecular dynamics we study equilibrium fluctuations and dynamical properties at the interface. Effect of bead size and interaction strength both on volume and volumetric fluctuations is manifested in mechanical properties, quantified here by bulk modulus, $K$. Tuning NP size and non-bonded interactions results in ~15% enhancement in $K$ by addition of a maximum of 1.5% v/v NP.

**INTRODUCTION**

Polymer matrix based nanocomposites have been attracting research interest due to their promise to improve mechanical properties such as higher stiffness, impact and tensile strength, as well as other rheological, viscoelestic and physical properties.[1-3] Since it is a big challenge to characterize the structure and properties, and manipulate the fabrication process, development of nanocomposites have mostly been empirical until recently.[4] Designed NPs are now routinely targeted for tuning polymer properties such as toughness, melt viscosity, and transient rubberlike behavior.[5]

In nanocomposites, interplay of different parameters such as temperature, filler size and shape, mixing ratio of matrix and filler, polymer type and size play significant roles in the enhanced materials properties.[6] In the last decade, computer modeling and simulations have had an increasing role in understanding and controlling the underlying mechanism of property changes and enhancements.[7-8] Fundamental issues such as molecular origin of reinforcement, rheological behavior for better processing and structure, and dynamics at the filler-matrix interface are the main purpose of these modeling and simulation efforts. Different computational methods, ranging from molecular to macro scale, such as molecular dynamics (MD),[9-11] Monte Carlo,[12-13] dissipative particle dynamics,[14] lattice Boltzmann,[15] equivalent-continuum [16] and finite element methods,[17] have been utilized for this purpose.

Inclusion of nanoparticles (NPs) introduces a large interfacial area, leading to a significant fraction of the chains to be labeled as interfacial polymer. These have properties substantially different from the bulk. Recent advances have made synthesis of NPs with size less than the polymer radius of gyration possible, and have invoked new questions. One interesting work has shown that tracer diffusion in polystyrene/fullerene or polystyrene/carbon nanotube nanocomposites displays a minimum at a critical NP loading.[18] These studies have paved the way for significant modification of polymer dynamics by the inclusion of vary small amounts of NPs (less than 1.5 volume %). Accordingly, the altered dynamics affect long time scales which in turn directly modify transport and mechanical properties.

To control and predict the properties of such nanocomposites, it is important to systematically modify interface interactions so as to study the properties of this region;[19] the behavior of polymers directly interacting with the attractive surface is also crucial.[20] MD simulations of model coarse-grained polymer-particle nanocomposites with repulsive, neutral and attractive interactions reveal that attractive systems display heterogeneous dynamics.[10]

In this manuscript, a systematic study of a model polymer matrix reinforced by a NP is presented. The NP is modeled at the atomistic detail which allows us to determine the dynamical and mechanical properties quantitatively. A NP that interacts with the polymer only through van der Waals interactions is incorporated in the bulk polymer and the effect of interaction strength is varied. Furthermore, the effect of the size of the NP is studied. We discuss the structure of the polymer at the interface as a function of NP size and NP-polymer interaction strength. Furthermore, we describe sub-nanosecond − nanosecond dynamics at the interface by using the escape times of polymeric chain units from the interface. Finally, we discuss the energetic and entropic origins of enhancement of mechanical properties and provide a recipe of optimal NP size/interaction strength for maximizing elastic moduli.

**MODELS AND METHODS**

**Molecular Model and Simulation Details.** NPs having radii in the range 3.2 − 7.2 Å are embedded in *cis*-1,4-polybutadine (PBD) chains to assess the effect of NPs on the properties of the polymer (figure 1). A total of 15 sets of MD simulations have been carried out (see Table 1) to test the effect of (i) cluster size and (ii) interaction strength between cluster atoms and polymer chains. Clusters consisting of $N$ atoms ($N$ = 10, 20, 30, 40, 70, 100 or 150), and 32 chains, each of 32 repeat units, are studied. Interaction strength is varied from very weak to very strong. To tune the strength, the well-depth for the van der Waals interaction between a NP atom and a united-atom carbon of the PBD chain, -$\varepsilon$, was varied in the range of 0.1 – 1.5 kcal/mol.

The PBD chains were simulated following the united atom model and force field parameterization of Gee and Boyd (see Table 2).[21] Using the same force field, PBD chains of length higher than 200 repeat units have been shown to be entangled.[22] Thus, the systems studied here are below the entanglement limit. Pure PBD was also studied to provide baseline values of the properties studied (Table 1, control group). In each simulation, there are a total of 4096 united atoms of PBD. The results reported here have been obtained at 330 K, well-above the glass transition temperature which is approximately at 170 K for *cis*-1,4-polybutadiene of 55000 g/cm$^3$ molecular weight.[23]

The geometries of the NPs used in this work are displayed in figure 2 along with the total number of outside interactions available to the NP (i.e., "free bonds"). The latter is calculated as follows: Each NP atom may have a maximum of 12 interactions available assuming close packing. Thus, the number of free bonds is simply the difference between 12$N$ and the number of interactions between NP atoms. We find a linear dependence on $R^2$ which is expected of spherical objects. Thus, we will assume all NPs studied in this work to be sphere-like, although they do have edge and corner atoms in the structure. Note that the objects in figure 2 are displayed as accessible surface areas, obtained by rolling a sphere of radius 2 Å (approximate size of CH/CH$_2$ units) along the surface of the object. We then calculate the enclosed surface area and the volume of the NP and calculate the radii assuming perfect spheres. The radii obtained from the area and the volume differ by at most 0.1 Å and are listed in Table 1.

The interaction parameters of the atoms constituting NPs are selected to be similar to that of silicon. Thus, the atomic mass of a NP atom is 32 g/mol, the well depth for non-bonded interactions between pairs of atoms occurs at 0.854 kcal/mol, and the van der Waals radius is $r_{NP}$ = 2 Å. The particle coordinates of the NP are generated as follows: Lowest energy configurations of atomic clusters containing a large range of atoms are listed in the Cambridge Cluster Database.[24] For a given NP size, those coordinates are rescaled to have radius $r_{NP}$. Pairs that are within $\sqrt[6]{2}r_{NP} = 2.24$ Å of each other are connected by springs with a spring constant of 150 kcal/mol. Thus, NP atoms are held together throughout the simulations, but there is some degree of flexibility inside the NP. NP atoms otherwise interact with each other and the rest of the polymeric chains via the Lennard-Jones 6-12 potential. We note that the intramolecular interactions are fixed throughout the simulations, and only the intermolecular interaction strength (well-depth) is varied. Note that the value of $\varepsilon$ = -0.3 kcal/mol corresponds to the typical interaction strength between a NP and a PBD unit, had the interactions been determined simply via the geometrical average.

In all simulations, the cutoff distance on non-bonded interactions are 10 Å, which are smoothed with a switching function set on at 8 Å. The time step is 2 fs and data are recorded at intervals of 1000 steps. Each simulation is carried out for 30 ns or longer (Table 1). Constructed boxes contain 4096 PBD united-atoms and the NP. Periodic boundary conditions are imposed. Calculations are carried out at constant temperature and pressure (NPT ensemble). The isotropic box sizes are in the range of 47-50 Å throughout the simulations.

To obtain well-equilibrated simulation boxes, we use the following procedure: (i) Construct the pure polymeric systems using the amorphous cell module of the Materials Studio suite of programs[25] at a density of 0.92 gr/cm$^3$; (ii) import the coordinates into the NAMD program[26] after deleting the hydrogen atoms; (iii) minimize at 300 K for 10000 steepest descents steps; (iv) carry out 50 ps NPT simulation at extremely low density (approximate 47 Å cubic sized chains immersed into 300 Å length box) and 300 K; (v) 150 ps NPT simulation at $10^6$ atm at 300 K in order to reduce the characteristic ratio of the chains to match previously reported values;[27] (vi) 1 ns NPT simulation at 1 atm and 430 K in order to relax the system. This is followed by data collection stage, for the duration indicated in Table 1. The first 5 ns of the simulations are discarded during the analyses. Apart from characteristic ratio, we also confirm equilibration by checking that the CH bond relaxation times agree with those obtained from NMR experiments.[28]

For the systems with the NP, we carry out further equilibration of the systems: For NPs with sizes from 10 to 70 atoms, the NP is directly embedded into the polymer matrix in a suitable void that has formed during the pure PBD simulations. This polymer-NP composite system is equilibrated for 2 ns at 1 atm pressure. For systems containing NP with 100 or 150 atoms, NP is first placed on the edge of an equilibrated polymer box, and the box dimension is extended to encase the NP. The resulting larger periodic box is simulated for 1 ns under 1000 atm pressure until a homogenous mixture is obtained and further equilibrated under 1 atm pressure and 430 K to reach density under ambient pressure. A sample equilibrated system is shown in figure 1. We then proceed with the data collection stage at the temperature of interest for the duration indicated in Table 1. Again, the first 5 ns of the simulations are discarded during the analyses.

**Residence times of chains near NP.** The residence times, $\tau_r$, of the chain units relate to the backbone dynamics and provide clues on the time scales at which chains lose memory of the presence of the NP. $\tau_r$ is the time required by an atom/molecule to escape from a given region [29-30] and is used to obtain information on the dynamical behavior of the polymer chains that are close to the surface of the NP. We have calculated $\tau_r$ by first marking, at a selected time origin, the PBD units in van der Waals contact with the NP; these are the units residing at a distance of 4 Å from the NP atom centers. We then monitor the decrease in the number of such PBD units as a function of time; even if a center leaves and comes back within the time frame of the observations, we still count the first passage of the center from the marked area. The resolution of the observations are 2 ps, the time interval of recorded snapshots from the MD trajectories. We repeat the calculations for a number of time origins throughout the recorded trajectories and we normalize the decays so that they have a value of $C(\tau_r) = 1$ at time 0. This is similar in spirit to the desorption times used to characterize the dynamics of polymeric chains on flat surfaces.[31]

The obtained decay curves are modeled by the superposition of a single-exponential term on the stretched-exponential model. The former captures the long-time tail while the latter models the initial decay. Thus, both the short- and long-time behavior of the $C(\tau_r)$ decays, as well as the crossover is well-reproduced:

$$C(\tau_r) = a \exp\left[-\left(\frac{t}{\tau_f}\right)^\beta\right] + (1-a)\exp\left[-\left(\frac{t}{\tau_s}\right)\right] = C_f + C_s \quad (1)$$

Here, $C_f$ and $C_s$ are the contributions to the overall correlations from the fast and slow processes, respectively; $\tau_f$ and $\tau_s$ are the corresponding characteristic times. The positive front factor, $0 \leq a \leq 1$, describes the relative contributions from processes at these separated timescales. $\beta$ is a dimensionless exponent carrying information on superposed single-exponential decays for the range $0 \leq \beta \leq 1$, following from its Laplace transform.[32] A time decay that may be interpreted via a stretched exponential function involves a dashpot with a time-dependent damping coefficient. The initial infinite decay rate of a stretched exponential function follows from its power-law descriptor [33] and is physically ascribed to a purely elastic response preceding a viscous decay, accounting for the elastic contribution. The characteristic times $\tau_f$ obtained from fits to a stretched exponential model are scaled with the $\beta$-exponent; for $\beta < 1$, they effectively extend into timescales slower than $\tau_f$ (see, e.g. figure 6 in reference[34] sample distribution of relaxation times for $\beta$=0.2 and 0.4).

To find the best curve fit to the decays using the four-parameter model in equation 1, we perform an exhaustive search in the space of permitted values of $a$, $\beta$, $\tau_f$, and $\tau_s$. $a$ and $\beta$ are incremented in units of 0.01 in the interval (0,1), $\tau_f$ is incremented in units of 2 ps in the interval (2,100), and $\tau_s$ is incremented in units of 50 ps in the interval (50,5000). The optimal solution is identified as the global least-squares error minimizer between the fitted and calculated $C(\tau_r)$ curves. To prevent excessive weighting of long-time tails at the expense of a poorly captured crossover region, we use the 1 ns portion of the curves to select the best-fitting parameters. Once different solutions are reached in this way, the final solution is verified to be the global least-squares error minimizer over the whole curve.

**RESULTS**

A recent work by the Rubinstein group has laid out the foundations of the mobility of nonsticky NPs in polymeric liquids.[35] Particles moving in a polymer matrix are mainly classified as those whose diameters *d* are (i) smaller than the correlation length of the polymer chains, i.e. the average distance between a monomer of one chain to a monomer on another chain, $\xi$; (ii) intermediate between $\xi$ and the tube diameter within which the polymers reptate; and (iii) larger than the tube diameter. The systems studied herein mainly represent (i) and (ii), both of whose terminal diffusion behavior is represented by the Stokes-Einstein relationship. In the former, the diffusion coefficient is inversely proportional to the polymer viscosity, $\eta$, while in the latter, it is inversely proportional to an effective viscosity, $\eta_{eff}$, which is rescaled as $\eta_{eff} = \eta (d/\xi)^2$. In this work, we further probe the regime where the intermediate-sized particles are also tuned to be "sticky" due to the strong interactions between its constituent atoms and the polymeric units. At the temperature and pressure of the simulations, we have calculated the correlation length to be $\xi = 12$ Å, which implies that the nanoparticles with $N \leq 40$ atoms are in regime (i), and the rest are in regime (ii). We also find that the long time slopes of mean square displacements of the NP are in accord with the two regimes (data not shown).

**Distribution of chains near nanoparticle surface – Effect of NP size**. The system is composed of two microphases: The NP and the vicinal chains that are within the range of its interactions constitute one phase; the PBD chains that do not directly "feel" the presence of the NP form the second microphase. The effective range of the NP may depend on both the geometry and the interaction strength. This range may be determined via calculating the radial distribution function (RDF) between the center-most atom of the NP (that closest to the center-of-mass) and the CH units of PBD, and is displayed in figure 3. The first coordination shell around the NP, constituting the CH units in direct contact with the NP, are marked by the gray dotted curves. This region displays either a single or double-peaked character, depending on the geometry of the NP. As the size of the NP is increased, the innermost atoms complete their 12 neighbors with other NP atoms and lose contact with the PBD chains. If a NP is considered as being formed by concentric layers, each layer may be partly or fully exposed, depending on the geometry of the NP. This factor determines the general shape of the RDF.

That an atom is exposed does not necessarily mean that it will complete its missing neighbors by nearby PBD chains. This is demonstrated for the case of the largest NP as an inset to figure 3. Therein, the NP atoms are shown as green spheres, while the parts of the PBD chains that are in van der Waals contact with any of the NP atoms is displayed in purple surface representation. Other nearby PBD units, those within 4 – 8 Å from the center of any NP atom, are also shown in transparent surface display. It is clear that there are depleted regions on the surface of the NP. This is due to the entropic cost of packing on a regular surface. The attractions between the NP and the PBD units do not fully compensate for this entropic factor. In fact, the density of PBD units per unit volume along the surface of the NP is constant for a given interaction strength. This is quantified by the integral of the first coordination shell (gray dotted parts of the curves in figure 3) which resides within $4.0 \pm 0.5$ Å of the surface atoms, yielding $4.9 \pm 0.1$ neighbors per unit volume for $\varepsilon = -0.3$ kcal/mol, regardless of NP size.

**Distribution of chains near nanoparticle surface – Effect of interaction strength**. The aforementioned entropic cost may be compensated energetically by introducing stronger interactions between the NP and the polymeric chains. In figure 4, we display the RDFs for the largest NP whereby the attractive interactions are tuned from weak ($\varepsilon = -0.1$ kcal/mol) to very strong ($\varepsilon = -1.5$ kcal/mol). For weak interactions, the distinction between the first and second coordination shells is slight, but the difference becomes more pronounced as the stickiness of the NP is increased: Not only does the first coordination shell get denser, but also the second shell draws nearer while getting denser. This results in the compaction of the PBD units residing along the NP surface as well as achieving complete coverage of the surface for $-\varepsilon > 0.75$ kcal/mol (inset to figure 4). Integrating the first coordination shell in figure 4, the number density of particles in the region increases from 4.6 to 5.7 neighbors per unit volume as the interactions are increased from the weakest to the strongest. In the next subsection we investigate how the distribution of chains near the NP surface translates into the dynamical features observed.

**Dynamics of chains near nanoparticle surface.** We observe in the RDFs that while the first coordination shell is significantly modified around the NP, the second coordination shell also displays organizational features different than bulk PBD.Then the questions arise: Does the NP and the chains in its first coordination shell act in unison leading to an effective NP radius that explains the main trends in all observations? Is the effect of the second coordination shell a minor perturbation, or does it contribute significantly to observed properties? Such effects would best manifest themselves in the dynamics within the region. We quantify it by the residence times, $\tau_r$, of the chain units within the first coordination shell of the NP (see Models and Methods for details).

The average decay curves of $\tau_r$, $C(\tau_r)$, are displayed in figures 5 and 6. The effect of increasing size displays separate character for larger NPs (figure 5a) and those having sizes $N = 10, 20$ and $30$ (figure 5b). On the other hand, the decays have a monotonous character for changing $\varepsilon$ values, displaying slower loss of memory of the NP with increasing interaction strength (figure 6a). Regardless, all decays exhibit two separate time scales. They are an initial decay on the tens of picoseconds, with a superposed slower nanosecond time scale. The initial decay (up to 100 ps) of the relaxation curves is well characterized by the stretched exponential form.

For all curves, we find that the stretch exponent $\beta$ is $0.50\pm0.05$ and the relevant characteristic time, $\tau_f$, is $50\pm20$ ps, regardless of the choice of $\varepsilon$ and NP size. On the other hand, the characteristic time for the slower component, $\tau_s$, increases from sub-nanosecond scales for weak interactions to over 4 ns for the strongest interactions (figure 6b). The effect of NP size on the slow component of chain residence times is more complex, and attributes two different regimes for the particle sizes studied here. The decays for $N \geq 40$ display the expected behavior of slowing down (figure 5a). The decays for $N = 10, 20$ and $30$ are actually slower that those of $N = 40$ (figure 5b).

We are thus in a position to interpret the physical meaning of $\tau_s$ and $\tau_f$. The fast decays are solely due to the motion of the PBD chains around the NP. These occur as a distribution of characteristic times resulting from the variety of the processes involved. Such relaxations are an inherent property of the polymeric units since neither the size of the NP nor the strength of the NP–PBD interactions affects $\tau_f$ and $\beta$.

On the other hand, the slow part of the decay is influenced by the motions of the NP whose timescale of motions are on the order of nanoseconds. The diffusion of a small NP is freely accommodated by the space between the chains for sizes below that of the correlation length of $\xi = 12$ Å. For these smaller NPs, $\tau_s$ describes the characteristic time scale of the NP diffusion. The PBD units "forget" the memory of the NP as it gradually leaves their vicinity. Since the radius of the $N = 40$ NP is 4.8 Å, and considering that the radius of a PBD unit is ca. 2 Å, the dimension of the NP reaches that of $\xi$ at this size. Thus, for $N = 40$ and larger, the NP has a waiting time until there is the right environment to facilitate its hopping to a neighboring site, its diffusion becoming delimited by the PBD chain motions.

Such hopping events require two conditions to be satisfied simultaneously: (i) During the dynamic motion of the polymeric units, an arrangement of the chains that open up a large enough space for the NP to pass should occur; (ii) the NP should have enough energy to detach itself from its current cavity and pass to the next one. The first condition has a threshold of size $\xi$ above which it becomes a less frequent event as the NP size increases. The second condition may be facilitated by increased temperature, but is harder to accomplish as the NP – PBD interactions are stronger. Increasing the interaction strength thus increases $\tau_s$, as displayed in figure 6b. Moreover, the second coordination shell significantly alters the dynamics, since $\tau_s$ continues to grow with increased interaction strength, even beyond $\varepsilon = -1.0$ kcal/mol where the region of van der Waals contact of the NP has reached maximum density (figure 3).

**Enhanced Mechanical Properties.** The above analyses show that the structure of the interface may be modified by NP cohesiveness. At the continuum limit, i.e., with large inclusions in the polymer matrix, a simple rule of mixtures described by the Mori-Tanaka Model predicts the stiffness tensor.[36] However, as the inclusion size is decreased, the physicochemical properties of the interface must be taken into account for a correct prediction of observed moduli which display substantial dependence on inclusion size.[37]

The dynamics observed at the interface is complex and depends both on size and stickiness of the NP. This will inevitably reflect on the mechanical properties. Below, we set out to determine the extent of enhancement of the mechanical properties of a polymer by the addition of miniscule amounts of NPs. We focus on the bulk modulus since the Poisson's ratio of PBD is 0.5, enabling all other moduli to be predicted. We note that we also calculated the shear moduli independently, using the implementation of the work of Theodorou and Suter[38] by the Materials Studio Program and verified that the Poisson's ratio is not affected by the presence of the NP.

Addition of an NP into the polymeric melt results in non-ideal mixing with the total volume $V$ given by:

$$V = V^{PBD} + V^{NP} + V^{mix} \qquad (2)$$

where $V^{mix}$ contains all the non-ideal effects of both internal energetic and entropic origin: $V^{mix} = V_E^{mix} + V_S^{mix}$. The internal energetic component $V_E^{mix}$ depends on (i) the surface area of the interface, since each NP – PBD interaction brings about an additional reduction in volume and the number of interactions are proportional to the square of the NP radius, $R^2$ (recall figure 2), (ii) interaction strength $\varepsilon$. The entropic component originates from mixing entropy, with $[\ln \varphi + \ln(1-\varphi)]$ dependence, which, at low volume fractions such as those studied here, has the major contribution from the first term. Thus, the leading term in $V_S^{mix}$ is obtained from the derivative of $-TS^{mix}$, varying as $\varphi^{-1} \sim R^{-3}$.

The energetic and entropic terms are expected to have compensating effects at low $\varphi$, but the former will take over as NP size is increased ($R^2$ dependence) while the latter component approaches zero. The variation in volume with NP size is shown in figure 7a. At low $R$ there is no change in the total volume; the entropy and enthalpy compensate for the increased size due to the presence of the NP. At large $R$ the major contribution will come directly from the $V^{NP}$ term, whose $R^3$ dependence dominates. A cubic fit to the data is shown in the figure to guide the eye. Note that the presence of the NP reduces the overall volume with respect to pure PBD except at the highest NP size.

For increased interaction strength at fixed large NP, the $V^{mix}$ term in equation 2 has contribution only from energetics. Thus, a roughly linear decreasing trend is expected in this case, with the box volume shrinking further as $\varepsilon$ is strengthened. Note that the complete crowding of the NP surface is not a limiting factor for decreasing volume, since the number density in the second coordination shell is also affected (recall figure 4). The variation in volume with $\varepsilon$ is shown in figure 7b which also displays the expected linear decreasing trend, shown to guide the eye.

The bulk modulus, $K$, is defined by:

$$K = -V \left.\frac{\partial P}{\partial V}\right|_T \quad (3)$$

where $P$ is pressure, $V$ is volume, and partial derivative is at constant temperature. Using MD trajactories produced in the NPT ensemble, we calculate the bulk modulus via,

$$K = \frac{k_B T \langle V \rangle}{\langle V^2 \rangle - \langle V \rangle^2} \quad (4)$$

Since the modulus depends on the ratio of volume to volume fluctuations, the resulting behavior is more complicated than that of the volume:

$$K = -\frac{V^{PBD} + V^{NP} + V^{mix}_E + V^{mix}_S}{\dfrac{-V^{PBD}}{K_0} + \dfrac{\partial V^{mix}_E}{\partial P} + \dfrac{\partial V^{mix}_S}{\partial P}} \quad (5)$$

with $V^{PBD}$ and $K_0$ being the volume and the modulus of the pure PBD system, respectively, and the NP assumed incompressible. Using $\partial V/\partial P=(\partial V/\partial\varphi)(\partial\varphi/\partial P)$ and assuming Raoult's law is valid so that $(\partial\varphi/\partial P)$ is a negative constant with value on the order of the inverse pressure at the very low concentrations that we study, we have, at $\varphi \ll 1$,

$$K = \frac{V^{PBD} + \mathcal{O}(R^3) - \mathcal{O}(\varepsilon R^2) - \mathcal{O}(R^{-3})}{\dfrac{V^{PBD}}{K_0} - \mathcal{O}(\varepsilon R^3) - \mathcal{O}(R^{-6})} \qquad (6)$$

with the $\mathcal{O}(\cdot)$ representing the order of the $R$ dependence of each term. Thus, we expect an initial increase in the bulk modulus at low $R$ due to the interplay of entropic part of the volume of mixing and fluctuations of entropic origin; as $R \to 0$, $K = \left[V^{PBD} - \mathcal{O}(R^{-3})\right] / \left[\left(V^{PBD}/K_0\right) - \mathcal{O}(R^{-6})\right]$. As $R$ gets larger, both entropic terms lose their effect, and the energetic contribution takes over: $K \approx \left[V^{PBD} + \mathcal{O}(R^3) - \mathcal{O}(\varepsilon R^2)\right] / \left[\left(V^{PBD}/K_0\right) - \mathcal{O}(\varepsilon R^3)\right]$ and a decrease in the modulus will occur due to fluctuations in the interaction zone decreasing faster than the reduction in the volume in the same region. Moreover, for a system with large $R$, the modulus may further be increased by strengthening NP – PBD interactions (more negative $\varepsilon$) which amplifies the latter effect.

The results are displayed in figures 7c and 7d along with the best-fitting function of the form of equation 6. We find that the modulus may be increased by ca. 7% by increasing the NP size, and that it reaches a maximum at roughly 6 Å radius (nm-size particle). By strengthening the NP – PBD interactions, the increase is ca. 13 % in the range studied. Therefore, the addition of a maximum of 1.4 volume per cent of NP with cohesive, non-bonded interactions leads to a much larger enhancement in the modulus. The superlinear increase is due to the cohesiveness of the NP – polymer interface, but will not be observed for large NPs.

**CONCLUSIONS**

A thorough understanding of the interfacial structure and dynamics is necessary for achieving designed-in properties in complex materials. The time scales operating at the interface directly influence the response to perturbations in a wide range of frequencies. The dynamics, in turn, is affected by the structural heterogeneity at the interface. In this work, we systematically study the effect of NP size and interface interactions on structural, dynamical and mechanical behavior of the polymer matrix. We show that the interface structure is independent of the curvature of the NP, but may be significantly modified by strong non-bonded interactions (figures 3 and 4). These effects may be tuned so as to extend into the region not in direct contact with the NP (figure 4, second coordination shell).

The dynamics at the interface in the ps – ns range has two major contributions. On the tens of ps range, a distribution of relaxation times, characterized by the stretched exponent $\beta=0.5$, is observed, originating from the dynamics of polymeric chains. Neither the characteristic times, nor the shape of the distribution depends on NP size or NP – PBD interaction strength. On the ns time scale, the dynamics is directly determined by the presence of the NP. Two separate regimes occur: (i) NPs of size less than the correlation length of the chains, $\xi = 12$ Å, freely move in the intervening space. In this case, the slow component of the relaxations directly measure the NP free diffusion times and slow down with NP size. (ii) For larger NPs, the slow component measures the waiting time for the NP to jump to adjacent vacancies of similar size. This component also depends on the NP to gather enough energy to detach itself from its current position, its value increasing with more negative $\varepsilon$ values. In effect, the NP smoothly leaves the vicinity of the chains it directly contacts in (i), while it loses its neighbors abruptly in (ii).

We find that the effective NP size extends further than the region of van der Waals contact. Although the first coordination shell saturates as interaction strength between the NP and PBD units is increased to $-\varepsilon > 0.75$ kcal/mol, the nanosecond time scale motions contributing to PBD residence times continues to slow down. This manifests the significant role played on the dynamics of the system by the reorganization taking place in regions extending further than the NP.

These structural and dynamical effects are reflected in mechanical properties, quantified here by the bulk modulus $K$. There is initially a superlinear increase in $K$ with increasing NP size, for fixed NP – PBD interactions, originating in the relative contributions of entropic factors to volume of mixing and volumetric fluctuations. This effect is counteracted by the hindered fluctuations of the chains brought about by the presence of the NP. We find that the upper bound in $K$ may be extended by enhancing the NP – PBD interactions, increasing the density around the NP and further supressing the fluctuations.

The detailed atomistic approach of the current study paves the way for modeling, forecast and control of polymeric properties in the presence of nanoscopic particles with attractive, rough surfaces. The results will find applications in microrheology studies used for probing dynamics of complex fluids[39-40] and in efforts for enhancing material properties of polymers by well-dispersed NPs.[7, 18] In particular, it provides design principles for multiscale modeling approaches where physics-based coarse graining to bridge the different levels of modeling proves to be detrimental for quantitative predictions.[41]

**Table 1. MD simulations carried out in this study***

|  | $N$ | $R$ (Å) | $-\varepsilon$ (kcal/mol) | volume fraction ($\varphi$) | simulation length (ns) |
|---|---|---|---|---|---|
| control | 0 | -- | -- | -- | 30 |
| N series | 10 | 3.2 | 0.30 | 0.0012 | 80 |
|  | 20 | 3.9 | 0.30 | 0.0021 | 80 |
|  | 30 | 4.4 | 0.30 | 0.0031 | 80 |
|  | 40 | 4.8 | 0.30 | 0.0041 | 40 |
|  | 70 | 5.7 | 0.30 | 0.0068 | 40 |
|  | 100 | 6.3 | 0.30 | 0.0093 | 80 |
|  | 150 | 7.2 | 0.30 | 0.0136 | 80 |
| $\varepsilon_{32}$ series | 150 | 7.2 | 0.10 | 0.0135 | 30 |
|  | 150 | 7.2 | 0.20 | 0.0135 | 30 |
|  | 150 | 7.2 | 0.30 | 0.0136 | 80 |
|  | 150 | 7.2 | 0.50 | 0.0136 | 30 |
|  | 150 | 7.2 | 0.75 | 0.0137 | 30 |
|  | 150 | 7.2 | 1.0 | 0.0137 | 60 |
|  | 150 | 7.2 | 1.25 | 0.0137 | 30 |
|  | 150 | 7.2 | 1.5 | 0.0138 | 30 |

*All carried out at 330 K, 1 atm pressure

**Table 2. Force field parameters for *cis* 1,4 poybutadiene [21] and NP atoms**

| Stretching | $V_l = k_l(l-l_0)^2/2$ | $k_l$ (kcal/mol/Å²) | $l_o$ (Å) |
|---|---|---|---|
|  | $CH_2 - CH_2$ | 158.5 | 1.54 |
|  | $CH - CH$ | 183.8 | 1.5 |
|  | $CH_2 - CH$ | 246.9 | 1.34 |
| Bending | $V_\theta = k_\theta(\theta-\theta_0)^2/2$ | $k_\theta$ (kcal/mol/rad²) | $\theta$ (degrees) |
|  | $CH_2 - CH_2 - CH_2$ | 115 | 111.65 |
|  | $CH_2 - CH - CH_2$ | 89.4 | 125.89 |
| Torsional | $V_\varphi = \sum_{n=1}^{6} k_n[1-\cos(n\varphi)]/2$ | $k_n$ (kcal/mol) |  |
|  | $CH_2 - CH - CH - CH_2$ | (0, 24.2,0,0,0,0) |  |
|  | $CH - CH - CH_2 - CH_2$ | (1.033, -0.472, 0.554, 0.263, 0.346, 0.164) |  |
|  | $CH - CH_2 - CH_2 - CH$ | (0.888, -0.619, -3.639, 0.666, -0.247, 0.190) |  |
| Nonbonded | $V_{ij} = 4\varepsilon\left[\left(\sigma/r_{ij}\right)^{12} - \left(\sigma/r_{ij}\right)^{6}\right]$ | $-\varepsilon$ (kcal/mol) | $\sigma$ (Å) |
|  | $CH_2 \cdots CH_2$ | 0.0936 | 4.500 |
|  | $CH \cdots CH$ | 0.1000 | 3.800 |
|  | $CH_2 \cdots CH$ | 0.1015 | 4.257 |
|  | NP atom $\cdots$ NP atom | 0.854 | 4.000 |
|  | NP atom $\cdots$ CH | Variable (see Table 1) | 3.900 |
|  | NP atom $\cdots$ $CH_2$ | Variable (see Table 1) | 4.250 |

**Figure Captions**

**Figure 1.** Sample simulation box of a NP in *cis*-1,4-polybutadine; (left) system of 32 chains, each with 32 repeat units with the inclusion of NP with $N=150$ atoms; (right) same system with the chains packed in the box. PBD chains are shown as transparent surface representations (color online only).

**Figure 2.** Relationship between surface area of NP and its free "bonds". Free bonds are calculated assuming each NP atom has the propensity to make 12 contacts. For each NP, the structure is shown as the accessible surface.

**Figure 3.** NP size dependence ($\varepsilon = -0.3$ kcal/mol) of the RDFs between the center-most atom of the NP and CH units in PBD chains. Vertical dashed lines indicate the radius of cluster. Gray thick-dotted parts mark the first coordination shell region where polymer units are in van der Waals contact with NP. Snapshot of surface coverage of NP for $N = 150$ is displayed as inset (Pink surface representation: PBD chain units in van der Waals contact with the NP; gray transparent surface representation: chain units close to, but not in van der Waals contact with the NP. Color online only.)

**Figure 4.** Interaction strength dependence of the RDFs between the centermost atom of the NP to CH units in PBD chains. The darker shades of gray indicate increased interaction strength. Inset: Snapshots of surface covereage of NP for $\varepsilon = -0.3$ kcal/mol (left) and $\varepsilon = -1.5$ kcal/mol (right); color scheme same as in figure 3; color online only.

**Figure 5.** Relaxation curves of surface residencies of PBD units: Dependence on size for the fixed interaction strength $\varepsilon = -0.3$ kcal/mol. Fast processes have the same distribution and characteristic decay times in both (a) and (b), quantified by $\beta = 0.50\pm0.05$ and $\tau_f = 50\pm20$ ps. **(a)** NPs of size greater than the correlation length, $\xi$, of PBD chains. $N = 40$ constitutes the borderline case with effective size just exceeding $\xi$ whereby its decay may be described by only the stretched exponential part, i.e. $a = 1.0$ for this particular case. Thus, $\tau_s$ contribution for $N = 40$ cannot be detected since hopping occurs on the sub-100 ns timescale coinciding with $\tau_f$. For $N = 70$, 100 and 150, $\tau_s$ values are 0.8, 1.7 and 2.5 ns, respectively. (b) NPs of size less than $\xi$; also reproduced for comparison is $N =$

40 (dotted). For $N$ = 10-20 and 30, $\tau_s$ values are 0.8 and 1.5 ns, respectively.

**Figure 6.** (a) Relaxation curves of surface residencies of PBD units: Dependence on interaction strength (attactions increased from $\varepsilon$ = -0.1 to -1.5 kcal/mol) for the fixed NP size of 150 atoms. (b) Slow and fast relaxation time components of the curves in (a). Just as in figure 5, fast component is independent of $\varepsilon$, with a value of 50 ± 20 ps and $\beta$ = 0.5 in all cases. The slow component displays nearly 10-fold increase in the window studied. Contribution of slower component to the overall decay also gets more dominant with increasing $\varepsilon$; $a$ in equation 1 decreases from 0.7 to 0.5.

**Figure 7.** (a) Effect of NP size on simulation box volume. Only the largest NP offsets the negative volume of mixing term. (b) Effect of interaction strength on simulation box volume. As the interactions are made more cohesive, the mixing term becomes more negative. (c) Effect of NP size on bulk modulus. (d) Effect of interaction strength on bulk modulus. In all figures, the value of the pure PBD system is shown by the horizantal dashed line. The expected trends are shown by the best-fitting dashed lines in each case. See equation 6 for (c) and (d), and its numerator for (a) and (b). In (a) and (b) error bars are smaller than the symbols.

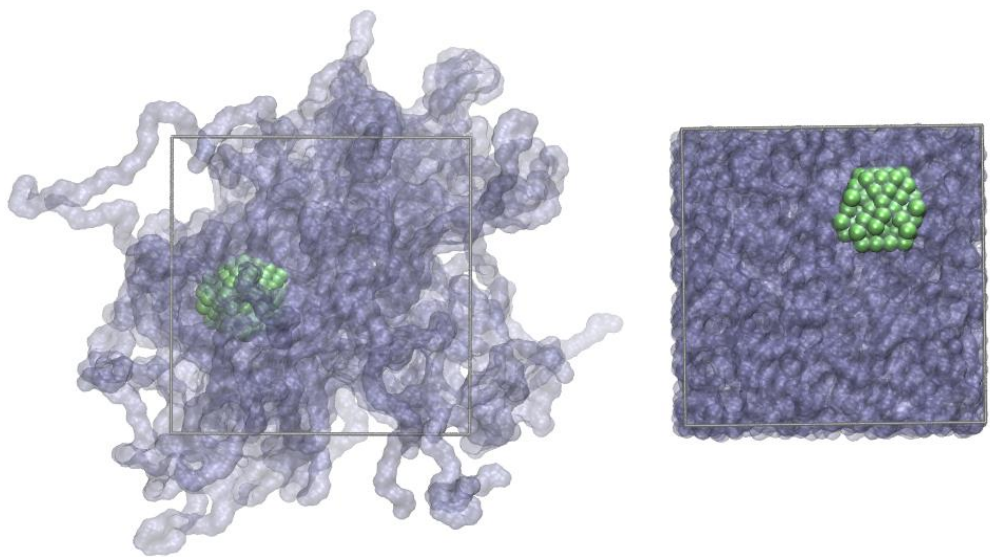

**Figure 1**

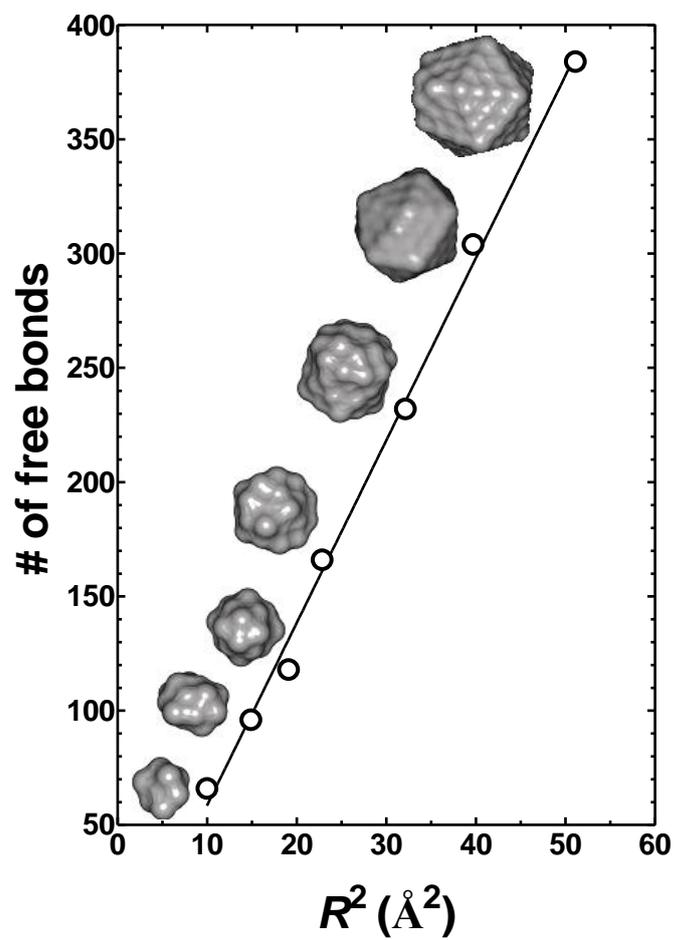

**Figure 2**

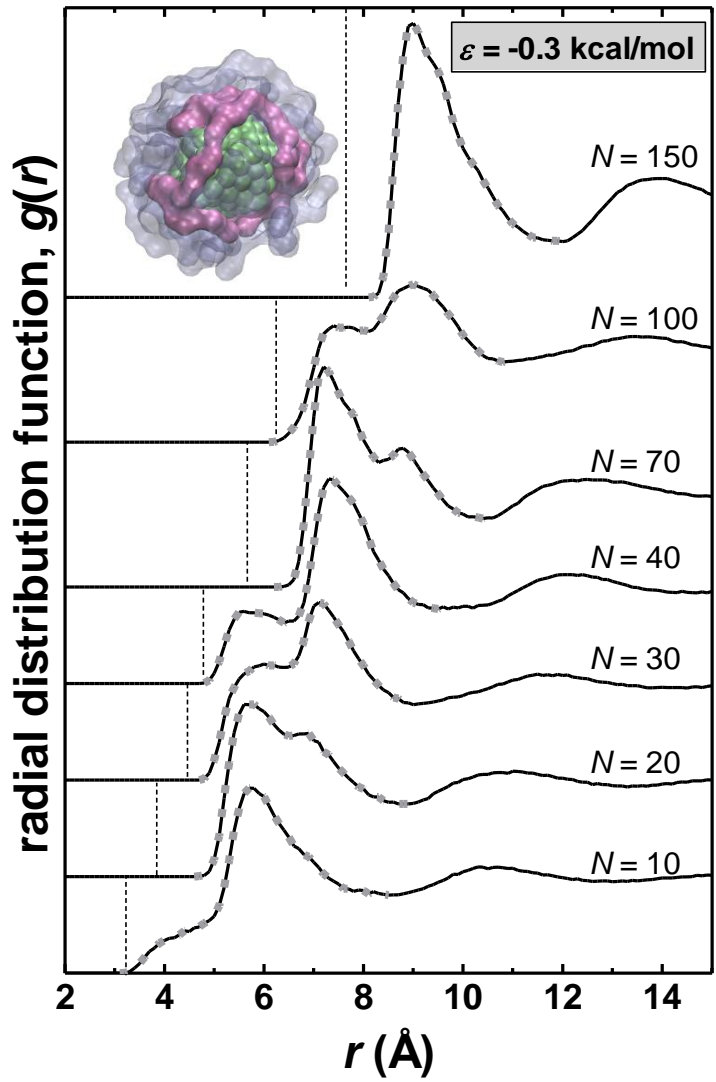

**Figure 3**

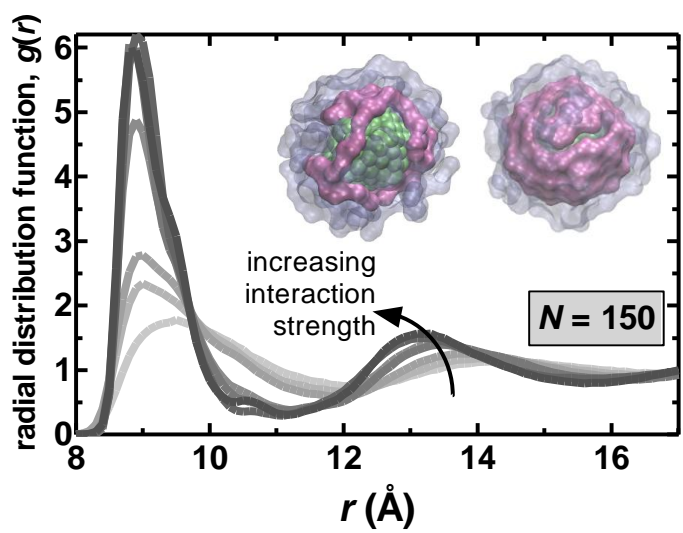

**Figure 4**

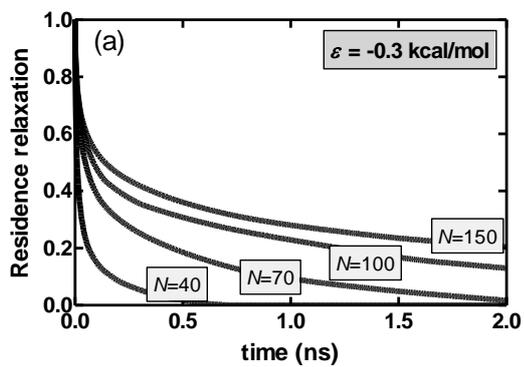
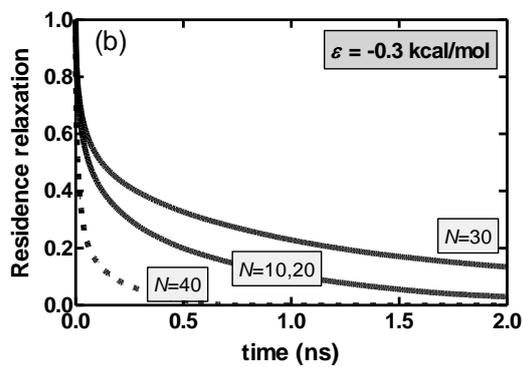

**Figure 5**

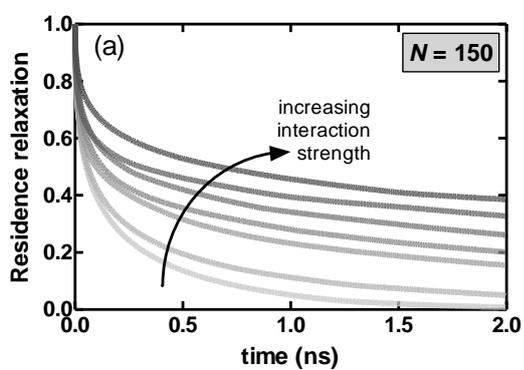
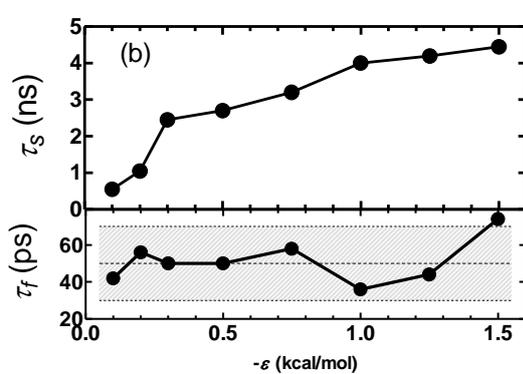

**Figure 6**

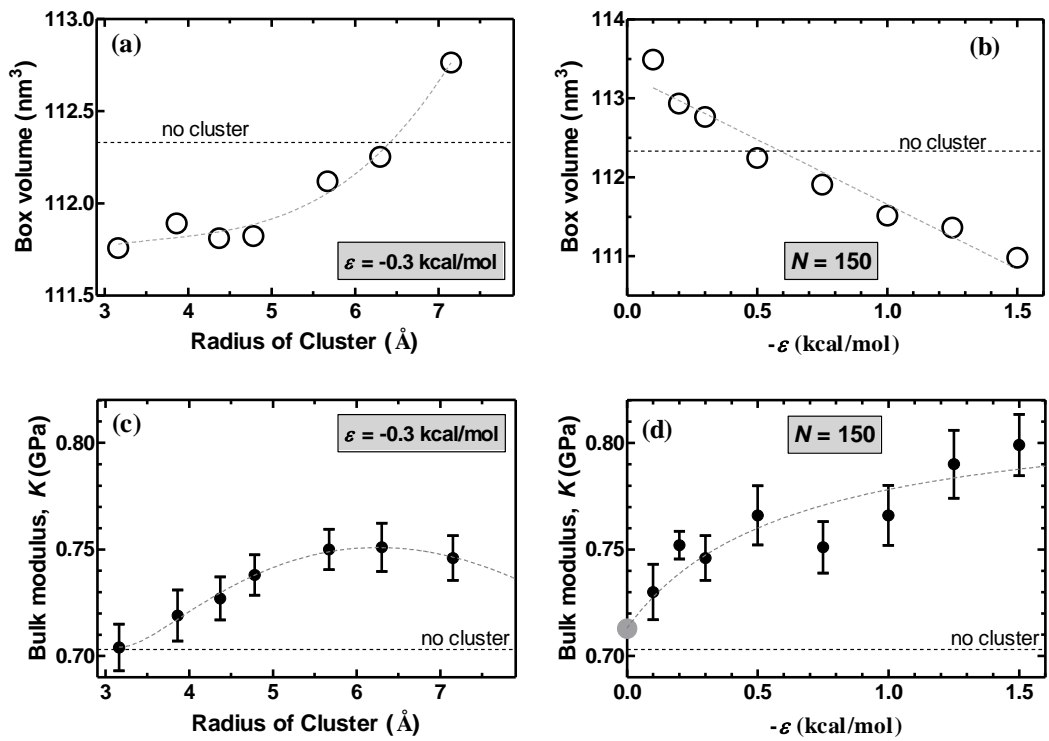

**Figure 7**